\documentclass[useAMS,usenatbib]{mn2e}
\usepackage{ram}
\def\kms{\mbox{$\rm km~s^{-1}$}}

\def\arcsec{^{\prime\prime}~}
\def\hb{\mbox{H$\beta~$}}
\def\hg{\mbox{H$\gamma~$}}
\def\hdelta{\mbox{H$\delta~$}}
\def\oi{[O\,{\sc{i}}]}
\def\o3{[O\,{\sc{iii}}]}
\def\s2{[S {\sc{ii}}]}
\def\n2{[N\,{\sc{ii}}]}
\def\oii{[O\,{\sc{ii}}]}
\def\ca2T{Ca\,{\sc{ii}} Triplet}

\begin{document}
\title[Star Formation in Bulgeless Galaxies]
{Star formation in bulgeless late type galaxies: clues to their evolution}
\author[Das et al.]{M.Das$^{1}$, C.~Sengupta$^{2, 3}$, S.~Ramya$^{1}$, K.~Misra$^{4}$ \\ 
1.~Indian Institute of Astrophysics, Koramangala, Bangalore 560034, India \\ 
2.~Instituto de Astrofísica de Andalucía (CSIC), Glorieta de Astronomía s/n, 18008 Granada, Spain\\
3.~Calar Alto Observatory, Centro Astronómico Hispano Alemán, C/ Jesús Durbán Remón, 2-2, 04004 Almeria, Spain\\
4.~Space Telescope Science Institute, 3700 San Martin Drive, Baltimore, MD 21218, USA} 

\date{Accepted.....; Received .....}


\maketitle


\begin{abstract}
We present GMRT 1280~MHz radio continuum  observations and follow-up optical studies 
of the disk and nuclear
star formation in a sample of low luminosity bulgeless galaxies. The main aim is to 
understand bulge formation and overall disk evolution in these late type 
galaxies. We detected
radio continuum from five of the twelve galaxies in our sample; the emission is mainly 
associated with disk star formation. Only two of the detected galaxies had extended radio 
emission; the others had patchy disk emission. In the former two galaxies, NGC~3445 and 
NGC~4027, the radio continuum is associated with star formation
triggered by tidal interactions with nearby companion galaxies. We did followup H$\alpha$ imaging
and nuclear spectroscopy of both galaxies using the Himalayan Chandra Telescope (HCT).
The H$\alpha$ emission is mainly associated with the strong spiral arms.
The nuclear spectra indicate ongoing nuclear star 
formation in NGC~3445 and NGC~4027 which maybe associated with nuclear star clusters. 
No obvious signs of AGN activity were detected. Although nearly bulgeless, 
both galaxies appear to have central oval distortions in the R band images; these could 
represent pseudobulges that may later evolve into large bulges. We thus conclude that 
tidal interactions are an important means of bulge formation and disk evolution in
bulgeless galaxies; without such triggers these galaxies appear to be low in star formation 
and overall disk evolution. 
\end{abstract}


\section{INTRODUCTION}

Bulgeless galaxies are an extreme class of late type spiral galaxiess (Scd to Sm) that have practically no bulge
and are nearly pure disk in morphology \citep{Boker.etal.2002}. Although these galaxies are not rare their formation and 
lack of evolution remains a puzzle both for Cold Dark Matter (CDM) theories of hierarchical galaxy formation \citep{White.Rees.1978},
and the secular evolution theories of galaxy disks \citep{Kormendy.Kennicutt.2004}. The former process leaves
clear signatures of merger history in the disks and the latter leads to the presence of oval distortions 
or disky bulges in the galaxy centers; neither of these features are seen in most bulgeless galaxies. 
One of the processes by which these disks could 
evolve is through interactions with nearby companion galaxies. The main aim of this work is to search for signs of such
evolution by mapping the star formation and nuclear activity in a sample of nearby bulgeless galaxies. This can 
help us better understand the transformation of these predominantly disk dominated galaxies into bulge dominated systems. 

Bulgeless galaxies show a wide variation in disk morphologies, ranging from the irregular dwarf galaxies to the nearly 
pure disk galaxies such as NGC~6503 \citep{kormendy.etal.2010}. They are late type spirals and have large, gas rich 
disks with low to moderate star formation rates. In some cases their galaxy disks may be optically bright as in NGC~5457 
(M~101), which has a relatively high star formation rate \citep{Boissier.etal.2007}. But most bulgeless galaxies have 
moderate to low surface brightness (LSB) disks, a good example being NGC~2552 which has a low luminosity disk 
\citep{deblok.etal.2005}. Being late type galaxies their dust content is often high 
(\citealt{popescu.etal.2002};\citealt{Ganda.etal.2009};\citealt{boselli.etal.2012}) and this 
can lead to significant optical extinction. Although bulgeless, small bars and oval distortions may be present as well as 
flocculent spiral arms that result in localized disk star formation. Such features are much lower in the more 
isolated bulgeless galaxies \citep[e.g.]{kautsch.2009}. The large scale spiral arms and  
bars that signify strong disk instabilities are not seen in most of bulgeless galaxies, especially the low luminosity 
(LL) ones.
This suggests that they may have dominant dark matter halos that stabilise their disks against the formation of bars,
spiral arms or oval perturbations. Indeed, for the LSB bulgeless galaxies this is
the case (\citealt{Zackrisson.etal.2006}; \citealt{Banerjee.etal.2010}) but this may also be true for the 
brighter bulgeless galaxies as well (\citealt{salucci.persic.1997}; \citealt{Kormendy.Freeman.2004}). Hence, only nearby 
interactions may be able to trigger disk instabilities that will lead to
star formation, nuclear gas inflow and central mass buildup in these galaxies. It is these processes that could result 
in bulge formation in such systems.

The formation and evolution of bulgeless galaxies in a $\Lambda$CDM universe is not well understood. Their angular momentum
and rotation velocities are not unusual for late type systems \citep{matthews.gallagher.2002}, but are too high to be
compatible with CDM models of galaxy formation \citep{Onghia.Burkert.2004}. Some models suggest that galactic outflows due to
supernovae and stellar winds can remove low angular momentum material from these disks leaving behind higher angular momentum
material in the disks (\citealt{Brook.etal.2011}; \citealt{Governato.etal.2010}). Or hot gas during the merging epoch is 
redistributed back to the disk via a galactic fountain \citep{brook.etal.2012}. However, this
process may not explain the nearly pure disk morphology for the larger bulgeless galaxies, especially those that are relatively
low in star formation actvity. Thus, although dynamically these galaxies represent the simplest form of disk galaxies, 
the physics behind their simple form is far more challenging to understand. 

Apart from the challenge of forming such pure disk galaxies, there is also the question of how they develop nuclear black 
holes and active galactic nuclei (AGN). Galaxy bulges and AGN are thought to grow
together in the nuclei of galaxies (e.g. \citealt{Silk.Rees.1998}; \citealt{Daddi.etal.2007}) and most AGN in our nearby Universe
are associated with bulges (e.g. \citealt{Heckman.etal.2004}). But suprisingly AGN have been detected in the nuclei of bulgeless
galaxies (\citealt{Filippenko.Ho.2003}; \citealt{Satyapal.etal.2007}; \citealt{Gliozzi.etal.2009}; \citealt{Satyapal.etal.2009}). 
Perhaps the most well studied example is NGC~4395 \citep{Filippenko.etal.1993}. The galaxy has a Seyfert~1 nucleus that 
has a black hole of mass $\sim10^{5}~M_{\odot}$ (\citealt{Peterson.etal.2005}; \citealt{Peterson.etal.2005}) which puts it in the
intermediate mass black holes (IMBH) domain. In fact most of the AGN detected in bulgeless galaxies have been found to be 
associated with IMBHs (\citealt{McAlpine.etal.2011}; \citealt{Shields.etal.2008}). It is not clear where these galaxies lie 
on the $M-\sigma$ correlation as they barely have a bulge; or how to explain their AGN activity which should be correlated with 
the formation of a bulge. Recent studies indicate that the SMBHs do not correlate with the disks or pseudobulges in galaxies
\citep{kormendy.etal.2011}, hence the disk may not be directly correlated with the growth of nuclear black holes in bulgeless galaxies. 
The AGN found in these galaxies are often buried inside nuclear star clusters which appear as bright cores in place of bulges 
(\citealt{Boker.etal.2002}; \citealt{Boker.etal.2004}; \citealt{Walcher.etal.2005}; \citealt{seth.etal.2008}; 
\citealt{Neumayer.etal.2011}).  
 
In this paper we investigate how star formation and nuclear activity in bulgeless galaxies can result in bulge formation disk evolution 
such galaxies. We focus only on LL bulgeless galaxies that may or may not have a bar, but do not have a prominent 
bulge. We use radio observations to map the star formation and search for compact nuclear emission in our sample of twelve 
galaxies. Radio emission has the advantage that it is not affected by dust obscuration and hence in this case may be a better 
tracer of the star formation.
For a deeper understanding we have followed up two interesting cases with optical spectroscopy and H$\alpha~$ imaging observations. 
In the following sections we discuss our sample, the observations and our results. We then summarise our findings and discuss their 
implications. 

\begin{table*}
 \centering
 \begin{minipage}{140mm}
  \caption{Galaxy Sample and Parameters}
  \begin{tabular}{@{}lccccccc@{}}
  \hline
   Galaxy & Other & Type & Distance & Position  & Velocity  & Optical & NVSS Radio \\
          &  Name &      &   (Mpc)  & (RA, DEC) &   (\kms)  &  Size   & Emission (mJy)\\
\hline
ESO~418-8 & PGC~013089 & SB(r)d & 14.9 & 03h31m30.6s, -30d12m48s & 1195 & 1.2' & 3.8 \\
UGC~4499  & PGC~024242 & SABdm & 11.5 & 08h37m41.5s, +51d39m09s & 691 & 1.99' & 2.0 \\
NGC~3346  & UGC~05842 & SB(rs)cd & 22  & 10h43m38.9s, +14d52m19 & 1260 & 2.7' & 13.8\\ 
NGC~3445  & UGC~06021 & SAB(s)m & 30.8 & 10h54m35.5s, +56d59m26s & 2069 & 1.6' & 23.6 \\
NGC~3782  & UGC~06618 & SAB(s)cd & 13.2 & 11h39m20.7s, +46d30m50s & 739 & 1.7' & 4.6 \\
NGC~3906  & UGC~06797 & SB(s)d  & 16.1 & 11h49m40.5s, +48d25m33s & 961 & 1.9' & 2.0 \\
NGC~4027  & PGC~037773 & SB(s)dm & 27.9 & 11h59m30.2s -19d15m55s & 1671 & 3.2' & 91.0 \\
NGC~4299  & UGC~07414 & SAB(s)dm  & 7.79 & 12h21m40.9s, +11d30m12s & 232 & 1.7' & 17.2 \\
NGC~4540  & UGC~07742 & SAB(rs)cd & 22.1 & 12h34m50.8s, +15d33m05s & 1286 & 1.9' & 3.1 \\
NGC~4701  & UGC~07975 & SA(s)cd & 14.5 & 12h49m11.6s, +03d23m19s & 721 & 2.8' & 18.5 \\
NGC~5584  & UGC~09201 & SAB(rs)cd & 26 & 14h22m23.8s, -00d23m16s & 1638 & 2.45' & 19.4 \\
NGC~5668  & UGC~09363 & SA(s)d & 25 & 14h33m24.3s +04d27m02s & 1582 & 3.3' & 23.5 \\
  \hline
\end{tabular}
\end{minipage}
\end{table*}

\section{Galaxy Sample}

Our sample consists of twelve late type LL spiral galaxies (Table~1) that appear to be bulgeless or have a minimal bulge
in their Hubble Space Telescope (HST) Wide Field and Planetary Camera 2 (WFPC2) images \citep{Boker.etal.2003}. 
The galaxies have a range of disk morphologies; some have  
nearly pure disks and some have progressively more distinct centers that may represent bulges in early stages 
of evolution. They all 
have compact stellar cores as indicated by their HST I band light profiles and are detected at some level in their
NRAO VLA Sky Survey (NVSS) radio maps. However, the resolution of NVSS is poor ($45\arcsec$) and hence does not show the detailed 
radio morphology. Only six galaxies have Very Large Array (VLA) Faint Images of the Radio Sky at Twenty Centimeters (FIRST) 
radio maps; UGC~4499, NGC~3445, NGC~3782, NGC~3906, NGC~4701 
and NGC~5584. FIRST has a smaller beam size ($5\arcsec$) and so the higher resolution is able to pick up the 
more intense emission that maybe associated with star formation. Of these six FIRST galaxies, only NGC~4701 shows radio 
emission and it is associated with the inner disk. In the following paragraph we briefly describe our sample galaxies.

\noindent
{\bf ESO~418-8, NGC~4299, NGC~4540 and UGC~4499~:~}All four galaxies have fairly featureless disks that do not 
support recent large scale star formation or spiral structure. Only knotty or very localised star forming regions are 
seen in the optical and UV maps \citep{gildepaz.etal.2007}. 
The NVSS maps reveal extended but relatively weak radio emission from the disks and is probably associated with 
the localized star forming regions.

\noindent
{\bf NGC~3346 and NGC~5668~:~}Both galaxies are relatively nearby and nearly face-on in orientation. NGC~3346 
has a small bar in the center that is associated with a fairly strong spiral structure. NGC~5668 has less
prominent spiral structure. The NVSS maps shows diffuse and extended radio
emission distributed over the inner disks of both galaxies.

\noindent
{\bf NGC3445~:~}Optical images suggest that this is a one armed spiral galaxy. The prominence of the southern spiral arm
gives the galaxy a lopsided appearance; only a closer inspection of the 2MASS K band image reveals the smaller 
northern arm. NGC~3445 forms part of a triplet 
of galaxies with two other galaxies, NGC~3440 and NGC~3458. Close to the southern arm and lying east of the nucleus of 
NGC~3445, there is a prominent star forming region that may have formed from a retrograde encounter with a neighbouring 
galaxy \citep{cao.etal.2007}. GALEX observations of the galaxy show that the star formation is extended over the disk 
\citep{smith.etal.2010} though much of it is associated with the spiral arms and the nucleus. The galaxy has a 
bright nucleus that hosts a nuclear star cluster \citep{seth.etal.2008} but no AGN.
The radio flux in the NVSS maps is large compared to other galaxies in our sample but due to the poor resolution of
NVSS, not much can be said about the radio morphology. This is one of the few galaxies in our sample that has been 
detected in CO emission \citep{Boker.etal.2003AA}; the molecular gas mass is relatively low and centrally concentrated.
The galaxy has a small oval bar in the center but bulge-disk separation was not possible in this galaxy, perhaps because 
the `bulge' is too small \citep{Baggett.etal.1998}.

\noindent
{\bf NGC~3782 and NGC~3906~:~}Both  galaxies have small bars in their centers but no strong spiral structure. Knots 
of star formation are distributed over their disks. The radio emission is diffuse and extended over the galaxy disks.  
NGC~3782 is fairly inclined ($i=60^{\circ}$) but  NGC~3906 is close to face-on in morphology.

\noindent
{\bf NGC~4027~:~}This galaxy, like NGC~3445, appears to be a one armed spiral galaxy in the optical images;
this is mainly due to the prominent northern spiral arm that gives it a lopsided appearance. The smaller southern spiral arm 
appears much more prominent in the HI images of the galaxy \citep{phookun.etal.1992} and extends down towards the smaller 
companion galaxy NGC~4027A. The net effect of the  tidal interaction between NGC~4027 and NGC~4027A has resulted in an
extended HI ring around NGC~4027. The star formation is distributed over the disk and 
and especially along the spiral arms of NGC~4027. The 1.4~GHz radio continuum flux for this galaxy is the highest in the 
sample (Table~2) and the 
emission is distributed over the disk and nuclear regions. The molecular gas mass of this galaxy is suprisingly low 
\citep{Young.etal.1995} and less than $6\times10^{6}$~M$_{\odot}$ \citep{das.etal.2005}; it is close to the detection limits 
in the FCRAO survey. This is also the only galaxy in our sample that 
has a very small bar that appears as an oval distortion in the center of the galaxy \citep{phookun.etal.1992}. Its bulge 
to disk size is very low (~1/25) \citep{Baggett.etal.1998} which suggests that it may be a pseudobulge or a bulge in 
early stages of formation.

\noindent
{\bf NGC~4701 and NGC~5584~:~}Both galaxies have a small bright nucleus and a flocculent spiral structure. Star formation 
is associated with the spiral pattern. The NVSS images show extended but diffuse radio emission from the disks  
and is probably associated with star forming regions over the spiral arms. NGC~4701 also shows some emission in the higher
resolution FIRST radio map. The emission is concentrated about the nucleus and the inner disk.

\section{Radio Observations}

We observed the galaxies in radio continuum at 1280~MHz using the Giant Meterwave Radio Telescope (GMRT) located 
near Pune, India (Ananthakrishnan \& Rao 2002). Observations were done during May, 2008. Nearby radio source were 
used for phase calibration. Each galaxy had a two hour scan; excluding time spent on nearby calibrators the average on source 
time was about 1.5. The data was obtained in the native "lta" format, converted to FITS format and then 
analysed using AIPS \footnote{Astronomical Image Processing System (AIPS) is distributed by NRAO which is a facility of 
NSF and operated under 
cooperative agreement by Associated Universities, Inc.}. Bad data was iteratively edited and calibrated on a 
single channel until satifactory gain solutions were obtained using standard tasks in AIPS. This was used to generate
bandpass solutions. The central 110 channels were bandpass calibrated and averaged to obtain the
continuum database. This was then imaged using IMAGR. Both natural and uniform weighted maps of the galaxies 
were made to obtain the extended structures and see if there is any compact emission associated with the nuclei 
(Figures~1-5). We also experimented with different UV tapers to bring out the extended, fainter radio emission.

\begin{table*}
 \centering
 \begin{minipage}{140mm}
  \caption{FIR-Radio ratio (q) and Star Formation Rate for Galaxies Detected in Radio Emission}
  \begin{tabular}{@{}lccccc@{}}
  \hline
   Galaxy & Far-infrared flux &  q     & Star Formation             & GMRT Radio Emission & Beam \\
     Name & ($W~m^{-2}$)      &  value & Rate ($M_{\odot}~yr^{-1}$) & at 1.28~GHz (mJy)   & Size ($^{\prime\prime}$) \\
\hline
NGC~3445  & 12.16 & 2.14 & 1.2 & 12.0~$\pm$~3.3 & 8 \\
NGC~3782  & 6.86  & 2.28 & 0.2 & 10.3~$\pm$~3.9  & 45 \\
NGC~4027  & 74.79 & 2.34 & 4.2 & 77.5~$\pm$~10.9 & 8 \\1
NGC~4299  & 18.75 & 2.46 & ... & 4.1~$\pm$~1.9 & 7 \\
NGC~5668  & 15.43 & 2.24 & ... & 4.0~$\pm$~2.2 & 39 \\
  \hline
\end{tabular}
\footnotetext[1]{FIR flux ($W~m^{-2}$) = $1.26\times10^{-14}[2.58~S_{60} + S_{100}]$ }
\footnotetext[2]{q = log[$\frac{FIR}{3.75\times10^{12}}$/S$_{1.4~GHz}$]}
\footnotetext[3]{Star formation rate derived from the total infra-red luminosities of the galaxies ($L_{TIR}$) 
listed in Bell (2003) by using the relation for SFR also listed in the paper i.e. 
SFR~(M$_{\odot}$~yr$^{-1}$)~=~1.72$\times$10$^{-10}L_{TIR}$(1 + $\sqrt{}\frac{10^9}{L_{TIR}}$).}
\footnotetext[4]{GMRT fluxes were derived from the natural weighted or low resolution maps. The errors were derived 
from the noise in the maps (in units of mJy/beam) and the beam size. } 
\noindent
\end{minipage}
\end{table*}

\section{H$\alpha$ Imaging and Optical Spectroscopy of NGC~3445 and NGC~4027}

We did H$\alpha$ imaging and optical spectroscopy of the two galaxies, NGC~3445 and NGC~4027, that showed
extended emission in the GMRT radio continuum maps. The observations were done with the 
2m Himalayan Chandra Telescope (HCT) which is 
located at the Indian Astronomical Observatory (IAO) at Hanle. The telescope is remotely controlled from 
the Center for Research and Education in Science and Technology (CREST) which is part of the Indian 
Institute of Astrophyiscs (IIA) in Bangalore. The spectrum of NGC~3445 was obtained on a cloudless night on 
17 March 2010 and
NGC~4027 was observed on a clear night on 19 January 2010. The spectra were obtained using a $11\arcmin\times1\farcs92$ 
slit (\#167l) in combination with a grism \#7 (blue region) and grism \#8 (red region) which covers the wavelength ranges 
3700--7600 \AA \ and 5500--9000 \AA \ with 
dispersions of 1.46 \AA \ pixel$^{-1}$ and 1.26 \AA \ pixel$^{-1}$, respectively. The spectral resolution is 
around $\sim 8.7$ \AA \ (398 \kms\ FWHM or $\sigma=169$ \kms\ at H$\alpha$) for grism \#7 and $\sim7$ \AA \ 
($\sigma = 136$ and 103 \kms\ at H$\alpha$ and \ca2T\ respectively) for grism \#8. The spectra were obtained after exposing both 
the galaxies for about 2400~s in grism \#7. The spectrum of NGC~4027 in grism \#8 was obtained with an exposure of 1800~s.
Due to the low signal to noise (S/N) ratio of the grism \#8 spectrum of NGC~3445, we did not use it for further analysis.
The slit was placed at the centre of the 
galaxy covering a central region of $\sim2''\times5''$ ($1''$ corresponds to 135 pc at a redshift of $\sim0.0055$).

Data reduction was carried out using the standard tasks available within IRAF \footnote{Image 
Reduction \& Analysis Facility Software distributed by National Optical Astronomy Observatories, which 
are operated by the Association of Universities for Research in Astronomy, Inc., under co-operative 
agreement with the National Science Foundation} which includes bias subtraction, extraction of one 
dimensional spectra, wavelength calibration using the ferrous argon lamp for grism \#7 and ferrous neon 
lamp for grism \#8. The wavelength calibrated spectra were flux calibrated using one of the spectroscopic 
standards of \cite{oke90} observed on the same night and then corrected for the redshifts of the galaxies.
The blue and the red flux calibrated spectra of NGC~4027 
were combined together with the help of {\it scombine} within {\it specred} package using a 
suitable scale factor estimated at the flat continuum portions of the overlapping part of the spectra. 
The flux calibrated spectra of NGC~3445 and NGC~4027 are shown in Figures~10 and 12. 

The H$\alpha$ images of NGC~3445 and NGC~4027 were obtained on 17 March 2010 from HCT. The H$\alpha$ filter installed 
inside the Himalaya Faint Object Spectrograph system (HFOSC) is centered around 6550 \AA~ and has a bandwidth of 100 \AA. 
Bessells $R$ filter (centered around 6400 
\AA with a BW of $\sim1600$ \AA) is used for continuum subtraction similar to the procedure described in 
\cite{waller1990}.  As discussed by \cite{james.2004}, for dark nights, the scaled $R$ band exposure gives
an excellent continuum subtraction. Each of the H$\alpha$ frames was observed for 10 minutes. Sky subtracted unit exposure 
 time $R$ frame was scaled and subtracted out from the H$\alpha$ that was also sky subracted and normalized with respect 
to exposure time. The final continuum subtracted H$\alpha$ frames for two galaxies NGC~3445 and NGC~4027 are 
shown in the Figures \ref{ha_n3445} and \ref{ha_n4027}, respectively. 

\begin{table*}
 \centering
 \begin{minipage}{140mm}
  \caption{H$\alpha$ luminosities and Star Formation Rates for Individual Regions in NGC~3445 and NGC~4027}
  \begin{tabular}{@{}lccccc@{}}
  \hline
   Galaxy & Region & H$\alpha$ Luminosity (10$^{41}$erg~s$^{-1}$) & Star Formation Rate ($M_{\odot}~yr^{-1}$) & Number of O stars \\
\hline
NGC~3445  & 1  & 1.62  & 1.29 & 1.2$\times$10$^{4}$ \\
          & 2  & 0.18  & 0.14 & 1.3$\times$10$^{3}$ \\
          & 3  & 0.17  & 0.14 & 1.3$\times$10$^{3}$ \\
          & 4  & 0.10  & 0.08 & 0.7$\times$10$^{3}$ \\
\hline
NGC~4027  & 1  & 0.73  & 0.58 & 5.4$\times$10$^{3}$ \\
          & 2  & 0.29  & 0.23 & 2.1$\times$10$^{3}$ \\
          & 3  & 0.74  & 0.59 & 5.4$\times$10$^{3}$ \\
          & 4  & 2.04  & 1.62 & 1.5$\times$10$^{4}$ \\
          & 5  & 0.58  & 0.46 & 4.2$\times$10$^{3}$ \\
          & 6  & 0.37  & 0.30 & 2.7$\times$10$^{3}$ \\
          & 7  & 0.91  & 0.72 & 6.7$\times$10$^{3}$ \\
          & 8  & 0.71  & 0.56 & 5.2$\times$10$^{3}$ \\ 
  \hline
\end{tabular}
\footnotetext[1]{Star formation rate ($M_{\odot}~yr^{-1}$) = L(H$\alpha$)/(1.26$\times$10$^{41}$erg~s$^{-1}$) \citep{Kennicutt.1998}} 
\footnotetext[2]{Approximate number of O stars determined from number of ionizing photons 
N$_{Lyc}$=7.34$\times$10$^{11}$L(H$\alpha$)~photons~s$^{-1}$ where we have assumed the rate of ionizing photons from one
O5 star to be ~10$^{49}$~photons~s$^{-1}$ \citep{ravindranath.prabhu.1998}.}
\noindent
\end{minipage}
\end{table*}

\section{Results} 

\subsection{Radio Observations}
We have detected radio emission from five of the twelve galaxies in our sample (Table~2). Although all twelve 
galaxies have radio emission in the NVSS images, only five were detected in our study. There are several reasons 
for this : 1.~Some of the sources have NVSS flux which is extended and diffuse disk emission. These sources
had low level flux (2 to 3 mJy ) in NVSS, which is close to NVSS noise value. With GMRT we confirmed that
they are non-detections e.g. ESO 418, NGC4499, NGC3906, NGC4540. 2.~Our GMRT observations were only two hours 
scans of each source. Hence, the UV coverage 
is not very high. Deeper observations are required to detect the more diffuse emission. Hence only emission with 
high flux density is detected. A good example is NGC~4027 where the 
GMRT and NVSS fluxes are close in value. 3.~In some cases, such as NGC~3782, the GMRT flux is much less than 
the NVSS flux because of confusion with a much brighter source nearby. Since the NVSS beam is large, 
this emission from a nearby source is picked up. 4.~In one galaxy, NGC~4701, the data quality is not good enough 
(too noisy) to detect the source even though the FIRST map shows some emission close to the nucleus. In some cases,
heavy flagging was needed for the central baselines; this along with incomplete UV coverage, led to non detections 
or flux loss.

In the following paragraphs we discuss our results in more detail. The GMRT flux is quite high for the galaxies 
NGC~3445 and NGC~4027 and the NVSS and GMRT fluxes are comparable. In these galaxies the radio emission 
is extended and associated with massive star formation along the spiral arms. Interestingly, both these 
galaxies are also tidally interacting with nearby companion galaxies.

\noindent
{\bf (i)~NGC~3782, NGC~4299, NGC~5668~:}~All three galaxies show only patchy radio emission associated with the disk 
(e.g. Figure~1; NGC~4299) and is due to localized star formation. Thus although the galaxies show
extended radio emission at 1.4~GHz in the NVSS maps, our higher resolution GMRT observations could only detect small pockets
of localized emission. This suggests that unlike NGC~3445 and NGC~4027, there is not much ongoing high mass star formation in 
these galaxies.

\noindent
{\bf (ii)~NGC~3445~:}~The radio emission mainly follows the prominent southern spiral arm (Figure~2) and to some extent 
the less prominent northern arm. It is due to thermal emission associated
with star formation triggered by tidal interactions with the companion galaxies NGC~3440 and NGC~3458 \citep{Smith.etal.2007}. 
The peak flux 
detected in our observations is 2~mJy/beam (where the beam is 8$^{\arcsec}$ and map noise is 0.15~mJy/beam) and
is coincident with the large star forming region south of the galactic nucleus.
The total radio emission that we detect at 1280~MHz with the 8$^{\arcsec}$ beam is 11.3~mJy which is about half of the the 
flux detected at 1420~MHz by NVSS which has a much larger beam (Table~2). In the high resolution or uniformly weighted map (Figure~3)
the emission breaks up into two peaks. The brighter peak is associated with the southern spiral arm and the other
lies along the northern spiral arm.  Suprisingly there is no emission associated with the nucleus even though this galaxy does 
have a prominent nuclear star cluster \citep{seth.etal.2008}. We calculated the star formation rate from the total
infra-red luminosity (TIR); its value is $\approx 1~M_{\odot}~yr^{-1}$ \citep{Bell.2003}. Although this is not high it is 
significant for a relatively low luminosity galaxy. NGC~3445 is also one of the few galaxies in our sample that
has molecular hydrogen gas \citep{Boker.etal.2003AA}; it is mainly confined to the inner parts of the galaxy disk and is probably
associated with the disk star formation.   

\noindent
{\bf (iii)~NGC~4027~:}~This galaxy shows the maximum radio emission in our sample (Figure~4). The net flux is $\sim88~$mJy and 
is mainly
distributed around the southern spiral arm. The emission north of the galaxy center is patchy and does not follow the spiral
pattern as closely as the southern arm. The  emission along the southern spiral arm is concentrated at two locations.
There is also some emission associated with the galaxy nucleus; it could be due to nuclear star formation or weak AGN activity.
In the high resolution map (Figure~5) most of the emission is associated with the southern arm which indicates that the more intense
star formation is located in this part of the galaxy. There is also emission associated with the galaxy nucleus; it may be due 
to the nuclear star cluster \citep{seth.etal.2008} or a weak AGN. We determined the star formation rate from the total infra-red 
luminosity and its value is $\approx 4~M_{\odot}~yr^{-1}$ \citep{Bell.2003}. The radio morphology clearly indicates that 
star formation 
activity has been triggereed by the tidal interaction with the smaller galaxy NGC~4027A that lies south of the galaxy and closely 
follows the large HI ring that envelopes both galaxies. Suprisingly, the northern spiral arm is more prominent in the optical images
than the southern arm. This could be due to the orientation of the galaxies during the interaction. Hence the northern spiral 
arm is comprised of older stars due to previous star formation activity whereas the southern arm represents ongoing star formation
in NGC~4027.

\subsection{H$\alpha$ imaging of NGC~3445, NGC~4027 and the Residual R Images}

\noindent
The H$\alpha$ emission maps of NGC~3445 and NGC~4027 are shown in Figures 6 and 7. The main emission regions are circled
in red. We have determined the H$\alpha$ flux for these knots and estimated the star formation rate for these regions using the 
empirical relation of Kennicutt (1983); the values are listed in Table~3. We have also obtained the H$\alpha$ subtracted R
band images of the galaxies and examined the core structure in both galaxies.

\noindent
{\bf NGC~3445~:~}The distribution of H$\alpha$ emission in NGC~3445 follows the UV emission quite closely as seen in the GALEX images 
(\citealt{gildepaz.etal.2005};\citealt{smith.etal.2010}). There are two knots of star formation concentrated along the 
spiral arms; one lies north of the galactic
nucleus and the second, which is much larger, lies southwest of the nucleus (Figure~6). The radio emission also 
closely follows the H$\alpha$ emission as seen in the overlay (Figure~8). We did not detect any H$\alpha$ 
emission from the nucleus; only emission from star forming regions in the tidally interacting arms is observed. The radio 
continuum peak at 1280 MHz coincides with the strong HII regions located at the western and northern regions. Table~3 shows the
H$\alpha$ luminosities for the more prominent star forming knots as well as their respective star formation rates (SFR). 
The SFR is less than 1~$M_{\odot}~yr^{-1}$ for most regions except the large star formation knot located in the 
southern arm which has a SFR of $\approx$1.5~M$_{\odot}~yr^{-1}$. 

\noindent
{\bf NGC~4027~:~}The H$\alpha$ emission observed in NGC~4027 (Figure~7) is much higher than that observed in NGC~3445 and the FIR 
luminosity (Table~2) is also higher. This is probably because the rate of newly forming stars is closely related to the tidal interaction 
with the companion galaxy and the tidal interaction in NGC~4027 \citep{phookun.etal.1992} appears to be stronger than that in NGC~3445. 
The H$\alpha$ emission in NGC~4027 closely follows the spiral arms but is more concentrated along the southern arm where there are 
two large knots of star formation; these are clearly seen in the radio map overlay (Figure~9). As in NGC~3445, the peak of the 
1280~MHz radio emission coincides with the strong HII region complexes in the spiral arms. Using  luminosity of each of the knots 
we have derived the star formation rates (Table~3); the number of O stars present in each of the knots is also given in the Table~3 
and lies in the range 10$^{2}$ to 10$^{4}$. There is also H$\alpha$ emission associated with the nucleus which could be due to 
nuclear star formation and possibly an AGN (see next subsection). 

\noindent
{\bf Residual R band Image~:~}The H$\alpha$ subtracted residual R band images are not very deep but give an idea of the disk structure 
in both galaxies (e.g. Figure~10, NGC~4027). We used the iraf task 
{\it ellipse} to fit ellipses and examine whether there are oval distortions in the galaxy centers. NGC~4027 has an oval distortion
of ellipticity approximately 0.6 and position angle $83^{\circ}$. The size of the oval distortion is $\approx15^{\arcsec}$ or 0.65~kpc.
The structure is disky rather than spherical and suggests that it may be a pseudobulge that has formed from the buckling of a much 
smaller bar and is in the process of growing into a bulge \citep{sellwood.wilkinson.1993}. The bulge-disk decomposition of NGC~4027 
yields a bulge scale length of $r_{e}\sim 1^{\arcsec}$ and disk ${\mu}_{e}\sim 16.2^{\arcsec}$ \citep{Baggett.etal.1998}. 
The bulge size is much smaller than our estimate using ellipse fitting; 
perhaps because their decomposition measures the central core and not the whole oval structure. The R image of NGC~3445 also has a 
small bulge in its center but it 
is not as elliptical as the incipient "bulge" in NGC~4027. The oval distortion in NGC~3445 appears to have an ellipticity less than 
0.5 and a position angle $45^{\circ}$; the total size is less than 0.5~kpc. The bulge is too small for bulge disk decomposition
\citep{Baggett.etal.1998}. The images also show that both galaxies have spiral arms that are associated with the tidal interactions 
with nearby galaxies. The spiral structure is sharper in NGC~4027 than NGC~3445; possibly because the tidal interaction is stronger in
NGC~4027 \citep{phookun.etal.1992}.    

\subsection{Optical spectroscopy of NGC~3445 and NGC~4027}

\noindent
{\bf (i)~NGC~3445~:}~ The optical spectrum of NGC~3445 is displayed in Figure \ref{f:n3445_spec_r2}. 
The fluxes of emission lines are tabulated in the Table \ref{t:n3445_emi}. 
A strong underlying stellar population in absorption is detected at \hb, \hg and \hdelta wavelengths. The composite light 
from NGC~3445 is decomposed using an underlying stellar spectrum, an Fe{\sc~ii} template and a power-law continuum using a
the Levenberg-Marquardt ${\chi}^2$ minimization technique. 
The procedure followed is discussed in detail in \cite{ramya2011}. The spectrum of NGC~3445, the decomposed components 
and the model subtracted final spectrum are all plotted in Figure \ref{f:n3445_sub_z_r2}. The age of the underlying 
stellar population is estimated to be in the range $500$ Myr to $1$ Gyr indicating that a recent episode of star formation has occurred 
in the galaxy. The galaxy is also experiencing current episodes of star formation as seen from the H$\alpha$ images. 
We searched for signatures of hidden AGN activity but no broad H$\alpha$ emission line was detected even after subtracting the 
underlying stellar continuum. NGC~3445 is classified as a galaxy with nuclear star clusters \citep{seth.etal.2008} rather than 
hosting an AGN or a composite kind of nucleus. We investigated where NGC~3445 lies on the Baldwin, Phillips \& Terlevich (BPT) 
diagram; the values of log(\o3~$\lambda$\,5007/\hb)$\sim0.18$ and log(\n2\,$\lambda$\,6584/H$\alpha$)$\sim-0.62$ 
lie in the region occupied by star forming galaxies. 
The fluxes of emission lines (see Table \ref{t:n3445_emi}) were used to calculate ionic abundances 
of each of the species of O, N, S using the task {\sc ionic} from the {\sc stsdas} package of {\sc iraf}. 
The electron temperature and density are $\sim11107$ K and $\sim26$ cm$^{-3}$, respectively. The elemental abundances are 
estimated following the prescription of \cite{pil2010}. The oxygen abundance 12~+~log($\frac{O}{H}$)~=~8.4 and nitrogen 
abundance is estimated to be 12~+~log($\frac{N}{H}$)~=~7.56. The metallicity is only slightly less than solar 
(12~+~log($\frac{O}{H}$)~=~8.67) metallicity in value.

\begin{table*}
 \centering
 \begin{minipage}{140mm}
  \caption{Emission line intensities, ionic abundances and elemental abundances of atomic species obtained from spectra of
 the galaxy NGC~3445. }\label{t:n3445_emi}
  \begin{tabular}{@{}cccccc@{}}
  \hline
  Species & Wavelength & Flux & EqW & Ionic & Abundance \\

          & $\lambda$ & $^a$ & \AA & $\frac{X^+}{H^+}$  & 12~+~log($\frac{X^+}{H^+}$) \\
  \hline
\oii   &   7325 & 7.08056 &  -0.809  &  1.053\,E-4  &  8.46 \\
\n2    &   6548 & 42.367  &  -2.815  &  1.881\,E-5  &  7.71  \\
H$\alpha$ & 6563 & 473.356 &  -32.12  &   ---        &  ---   \\
\n2    &   6584 & 109.445 &  -7.645  &  1.652\,E-5  &  7.65  \\
\s2    &   6717 & 84.1673 &  -6.068  &  2.568\,E-6  &  6.84  \\
\s2    &   6731 & 58.6366 &  -4.286  &  2.495\,E-6  &  6.83  \\
He{\sc~i}    &   5876 & 17.9532 &  -1.052  &  ---       &   ---  \\
\hb    &   4861 & 100     &  -5.444  &  ---         &  ---  \\ 
\o3    &   4959 & 56.5241 &  -3.072  &  4.117\,E-5  &   8.05 \\
\o3    &   5007 & 158.82  &  -8.831  &  4.003\,E-5  &   8.04 \\
\hg    &   4340 & 23.4057 &  -1.349  &  ---         &   ---  \\
\oii   &   3727 & 343.514 &  -23.39  &  0.819\,E-4  &   8.35 \\
\\
 \multicolumn{4}{c}{F(H$\beta$) erg~cm$^{-2}$~s$^{-1}$} & \multicolumn{2}{c}{$1.97\times10^{-14}$} \\
 \multicolumn{4}{c}{12+log($\frac{O}{H}$)} & \multicolumn{2}{c}{8.40$^b$} \\
 \multicolumn{4}{c}{12+log($\frac{N}{H}$)} & \multicolumn{2}{c}{7.56$^b$} \\
  \hline
\end{tabular}
\footnotetext[1]{Fluxes of all species are normalised with respect to \hb flux and errors on the fluxes are 
in the range $5-10$\%.}
\footnotetext[2]{The oxygen and nitrogen abundances are estimated using the new calibration of \cite{pil2010}.}
\noindent
\end{minipage}
\end{table*}

\noindent
{\bf (ii)~NGC~4027~:}~ The spectrum of NGC~4027 is displayed in Figure \ref{f:n4027_spec_r2}. It is interesting to note
that the spectrum does not show any forbidden lines of Oxygen like \oii\,$\lambda$\,3727, \o3\,$\lambda\lambda$\,4959,5007 or even 
\oi\,$\lambda$\,6300. The non-detection of oxygen lines but presence of \n2\,$\lambda\lambda$\,6548,6584 and 
\s2\,$\lambda$~6717,6731 is surprising. It could be due to dust \citep{roussel.etal.2001} which may have used up most of the 
oxygen during the formation of dense clouds. Also, the Balmer decrement (H$\alpha$/H$\beta$) is approximately 6, 
which is much greater 
than the theoretically expected value of 3. Table~5 displays the emission line fluxes, equivalent widths, 
and ionic abundances. In comparison with NGC~3445, the nitrogen and sulphur ionic abundances of NGC~4027 are higher by about 
0.6 dex and 0.1 dex, respectively, which hints towards higher oxygen abundance and metallicity than NGC~3445. 
The stellar decomposition was 
carried out and the residual spectrum, Figure \ref{f:n4027_sub_z_r2}, neither shows any oxygen line nor show any broad 
H$\alpha$ emission confirming that this galaxy hosts a nuclear star cluster and not an AGN \citep{seth.etal.2008}. 
The age of the underlying stellar population as estimated from the best fit starburst99 template is $\sim1$ Gyr.
Using the emission line ratio of \s2 $\lambda$\,6717/6731, the electron density is estimated to be $273$ cm$^{-3}$. 

\begin{table*}
 \centering
 \begin{minipage}{140mm}
  \caption{Emission line intensities, ionic abundances and elemental abundances of atomic species obtained from spectra of
 the galaxy NGC~4027.  }\label{n4027_emi}
  \begin{tabular}{@{}cccccc@{}}
  \hline
  Species & Wavelength & Flux & EqW & Ionic &  Abundance \\

          & $\lambda$ & $^a$ & \AA & $\frac{X^+}{H^+}$  & 12~+~log($\frac{X^+}{H^+}$) \\
  \hline
\oii &   7325 & --- &  ---  &  ---  &  --- \\
\n2  &  6548  &  108.77	&   -4.65   &	6.29\,E-5   &    8.23 \\
H$\alpha$  &  6563  &  671.07	&   -29.22  &	  ---	     &    --- \\
\n2  &  6584  &  304.19	&   -13.58  &	5.98\,E-5   &    8.21 \\
\s2  &  6717  &  88.20	&   -3.60   &	3.91\,E-6   &    7.03 \\
\s2  &  6731  &  74.98	&   -3.06   &	3.91\,E-6   &    7.03 \\
He{\sc~i}  &  5876  &  33.66	&   -1.20   &	  ---	     &    --- \\
\hb  &  4861  &  100	&   -3.71   &	  ---	     &    --- \\
\o3    &   4959 & --- &  ---  &  ---  &   --- \\
\o3    &   5007 & ---  & ---  &  ---  &   --- \\
\hg    &   4340 & --- &  ---  &  ---  &   ---  \\
\oii   &   3727 & --- &  ---  &  ---  &   --- \\
\\
 \multicolumn{4}{c}{F(H$\beta$) erg~cm$^{-2}$~s$^{-1}$} & \multicolumn{2}{c}{$1.37\times10^{-14}$} \\

  \hline
\end{tabular}
\footnotetext[1]{Fluxes of all species are normalised with respect to \hb flux and errors on the fluxes are 
in the range $5-10$\% because of poor signal-to-noise of the optical spectrum.}
\footnotetext[2]{Since oxygen lines are undetected in the spectrum of NGC~4027, it was not possible to estimate the 
oxygen and nitrogen abundances.}
\noindent
\end{minipage}
\end{table*}

\section{Discussion}

\noindent
{\bf (i)~}Star Formation in Bulgeless Galaxies~:~Our radio observations as well as general observations in the literature suggest 
that bulgeless galaxies are generally poorly evolved systems that have low star formation rates (SFR) unless interacting with nearby 
galaxies \citep{kautsch.2009}. In our study all the galaxies are rich in HI \citep{watson.etal.2011} and hence have the capacity to 
support star formation, but most are fairly quiescent. We compared our galaxies with other pure disk galaxies that have been studied 
in the literature; such as the relatively low luminosity galay NGC~6503 and the brighter galaxy M101. We estimated their SFRs 
using the total infrared (TIR) luminosities
\citep{Bell.2003} as we did for our sample galaxies in Table~2. NGC~6503, which has a low SFR of $0.22~M_{\odot}~yr^{-1}$, is close to the 
local void \citep{karachentsev.etal.2003} and is hence fairly isolated. Its SFR is similar to NGC~3782 which shows only patchy star 
formation. M101 (NGC~5457), however, is a fairly bright star forming spiral and has a 
SFR~$4.1~M_{\odot}~yr^{-1}$; it is part of a group of mainly dwarf galaxies \citep{kinney.etal.1993}. Its SFR is similar to NGC~4027 but 
larger than NGC~3445; the star formation is probably triggered by interactions with nearby galaxies in the group. These two examples also
suggest that bulgeless galaxies are generally low in star formation unless interacting with nearby galaxies. This raises
the possibilty that their disks are more dark matter dominated than regular spirals and hence need a strong trigger, such as an 
interaction with a close companion galaxy, to start forming stars.

\noindent
{\bf (ii)~}Bulge growth in disk dominated galaxies~:~The main result of this work is that we have found two cases of ongoing 
bulge formation in two low luminosity bulgeless galaxies, NGC~3445 and NGC~4027 in which bulge formation has been triggered by 
close tidal encounters with nearby galaxies. Both galaxies are dark matter dominated systems since they
are late type spirals with low luminosity disks \citep{salucci.persic.1997}. Such systems are not easily perturbed to
form bars and spiral arms. However, interaction induced disk activity can result in enhanced star formation that can eventually 
lead to the build-up of bulges in the galaxy disks \citep{kormendy.fisher.2008}. This process may be ongoing in NGC~3445 where
a small bright core or bulge maybe developing. Alternatively, tidal interaction can trigger bar formation 
and this can lead to bulge growth through the formation of a pseudobulge. This may have 
happened in the case of NGC~4027 where the small oval distortion that we see in the R band image may have started out as a small bar 
instabilty; this then buckled to form an oval distortion 
or pseudobulge (\citealt{combes.etal.1990};\citealt{raha.etal.1991};\citealt{Martinez-Valpuesta.etal.2004}). Another example of 
interaction induced activity in a LL galaxy is NGC~4625. This galaxy is 
interacting with NGC~4618 and perhaps merging with 
NGC~4625A \citep{gildepaz.etal.2005}; the close interaction has triggered star formation in the extended low surface brightness 
(LSB) disk of NGC~4625. Thus, close 
tidal encounters are an important mechanism for the evolution of halo dominated systems such as LSB galaxies or low luminosity bulgeless 
galaxies such as those in our study. 

\noindent
{\bf (iii)~}AGN-bulge evolution~:~The well known black hole mass and bulge velocity relation 
(\citealt{ferrarese.merritt.2000}; \citealt{gebhardt.etal.2000}; \citealt{gultekin.etal.2009}) is difficult to understand in the 
context of bulgeless galaxies, since it implies that black hole growth is very difficult without a bulge. However, there is a 
significant number of 
bulgeless galaxies that show AGN activity and have intermediate mass black holes (IMBH), for example NGC~3367 and NGC~4536 
\citep{McAlpine.etal.2011}. Kormendy et al. (2011) have suggested that there are two ways of forming nuclear black holes; one driven by
mergers and gas infall; it leads to the formation of classical bulges and is often associated with SMBHs in massive galaxies. 
The second is through secular evolution processes the galaxy disks; this includes the formation of bars and pseudobulges. This latter 
process leads to the formation of smaller black holes or IMBHs in the centers of galaxies. In our study, NGC~3445 may be in the process
of forming a bulge through the first process, i.e. gas infall and central star formation whereas NGC~4027 may be an example
of the second mode of bulge growth. In NGC~4027 the formation of spiral arms and the growth of a 
central bar has been triggered by close tidal interactions with the nearby galaxy NGC~4027~A, leading to the formation of a pseudobulge 
(Figure~10) and weak AGN activity. Although neither NGC~3445 or NGC~4027 show optical emission lines characteristic of AGN, the AGN 
activity may 
star later. Studies indicate that the nuclear black holes begin to accrete and show AGN activity only after the initial starbursting phase 
is over \citep{Cen.2011}. The black holes formed in this manner, however, through pseudobulge evolution may not lie along the $M-\sigma$ 
relation \citep{kormendy.etal.2011}.  

\section{Conclusion}

\noindent
{\bf (i)~}We have detected radio emission at 1280~MHz from five of the twelve bulgeless galaxies in our sample. Of these five galaxies,
two galaxies, NGC~3445 and NGC~4027, show strong radio emission associated with their disks. Both galaxies are 
closely interacting with nearby companion galaxies and the radio emission is mainly due to the tidally triggered disk star formation.
NGC~4027 also shows some nuclear radio emission. In the 
remaining three galaxies, the emission was patchy and due to localized star forming regions in their disks.

\noindent
{\bf (ii)~}Both NGC~3445 and NGC~4027 have H$\alpha$ disk emission associated with star formation along their spiral arms. The radio 
emission closely follows the H$\alpha$ emission. The nuclear optical spectroscopy suggests that there is ongoing nuclear star formation 
in both NGC~3445 and NGC~4027; it is possibly associated with the formation of nuclear star clusters. No AGN emission lines were detected 
in the optical spectra of either galaxy.

\noindent
{\bf (iii)~}The optical R band images of both NGC~3445 and NGC~4027 show concentrations of mass in their nuclear regions. This may 
represent small bulges in early stages of formation. The close tidal interactions has led to the formation of extended spiral structure
and star formation which contribute to the build-up of nuclear mass. In NGC~4027 there is an oval distortion that may represent a 
pseudobulge. Thus close tidal interactions, rather than internal processes, may be important for bulge formation and disk evolution
in low luminosity, bulgeless galaxies. 

\section*{Acknowledgments}

We thank the GMRT staff for help in the observations. 
The GMRT is operated by the National Center for Radio Astrophysics of
the Tata Institute of Fundamental Research. We also thank the staff of HCT, IAO for the support during the observations.
This publication makes use of data products from the
Two Micron All Sky Survey, which is a joint project of the University of Massachusetts and the
Infrared Processing and Analysis Center/California Institute of Technology, funded by the
National Aeronautics and Space Administration and the National Science Foundation.
This research has made use of the NASA/IPAC Extragalactic Database (NED) which is operated by the
Jet Propulsion Laboratory, California Institute of Technology, under contract with the
National Aeronautics and Space Administration.

\bibliographystyle{mn2e}
\bibliography{mdas}

\begin{figure}
\includegraphics[width=80mm,height=80mm,angle=-90]{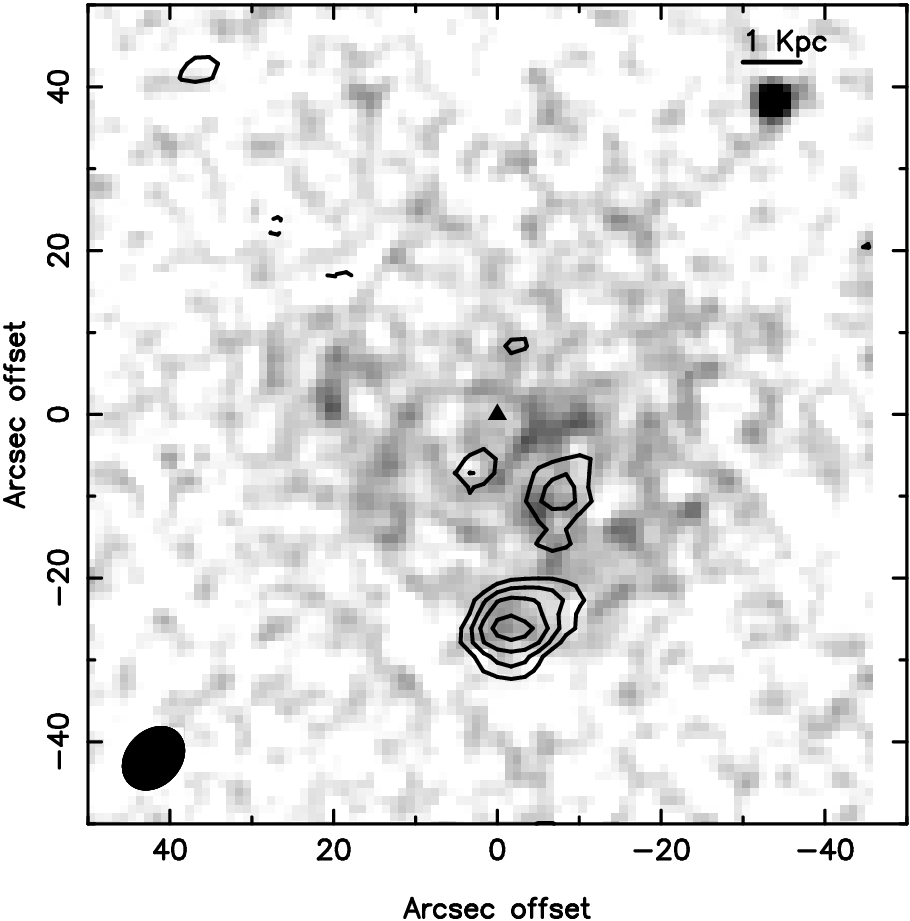}
\caption{Figure shows the contours of the GMRT 1280~MHz radio
continuum map of NGC~4299 superimposed on the 2MASS near-IR image of the galaxy. The peak radio flux
is $1.4~mJy~beam^{-1}$ and the $beam\sim7.3^{\prime\prime}$. The contours are 2, 3, 4, 5, 6 times the
noise level which is $0.25~mJy~beam^{-1}$. The galaxy center is marked with a filled triangle.
Note that the emission is offset from the galaxy center and lies mainly south of the nucleus in the disk.
}
\end{figure}

\begin{figure}
\includegraphics[width=80mm,height=80mm,angle=-90]{n3445.opt.radio.rsubmit.ps}
\caption{Figure shows the contours of the natural weighted GMRT 1280~MHz radio
continuum map of NGC~3445 superimposed on the 2MASS near-IR image of the galaxy. The peak radio flux
is $2~mJy~beam^{-1}$ and the $beam\sim8^{\prime\prime}$. The contours are 4, 6, 8, 10, 12 times the
noise level which is $0.15~mJy~beam^{-1}$. The galaxy center is marked with a filled triangle.
Note that most of the emission is offset from the center of the galaxy and lies mainly east of the 
nucleus.}
\end{figure}

\begin{figure}
\includegraphics[width=80mm,height=80mm,angle=-90]{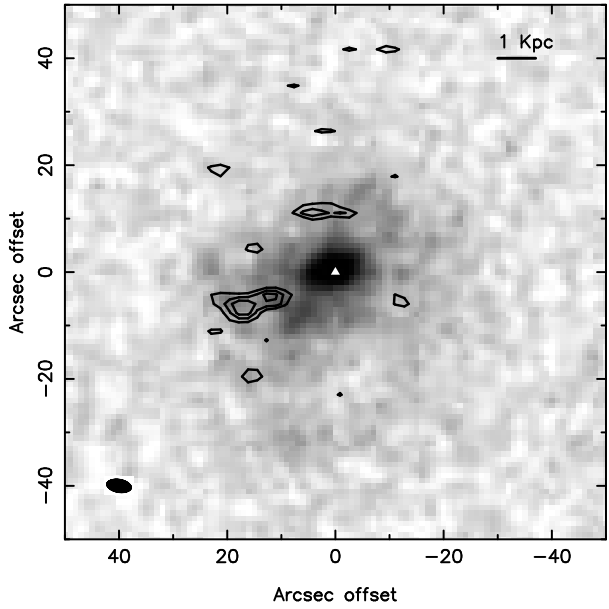}
\caption{Figure shows the contours of the uniformly weighted (or high resolution) 1280~MHz radio
continuum map of NGC~3445 superimposed on the 2MASS near-IR image of the galaxy. The peak radio flux
is $2~mJy~beam^{-1}$ and the $beam\sim5.5^{\prime\prime}\times 3.3^{\prime\prime}$. The 
contours are 0.6, 0.8, 1~mJy~$beam^{-1}$; this is 4, 5.7 and 7 times the
noise level which is $0.14~mJy~beam^{-1}$. The high density emission is offset from the galaxy center 
and lies mainly east and north of the nucleus.}
\end{figure}

\begin{figure}
\includegraphics[width=80mm,height=80mm,angle=-90]{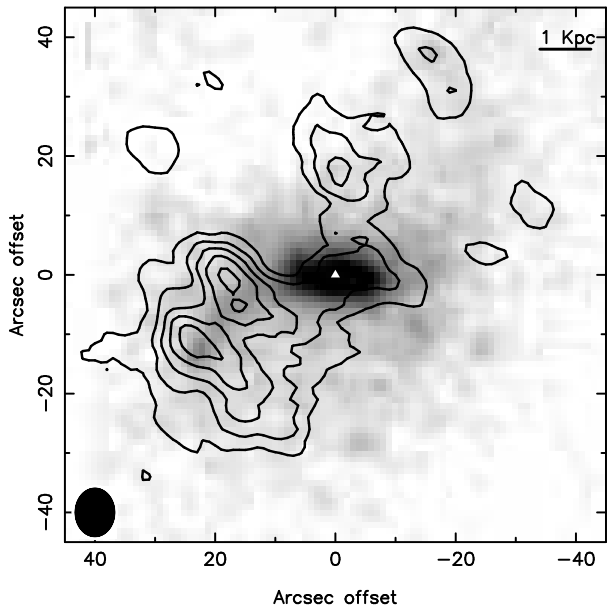}
\caption{Figure shows the contours of the natural weighted GMRT 1280~MHz radio
continuum map of NGC~4027 superimposed on the 2MASS near-IR image of the galaxy. The peak radio flux
is $\sim3.5~mJy~beam^{-1}$ where $beam\sim8^{\prime\prime}$; it is located in the disk and not the nucleus. 
The contours are 8, 10, 12, 14, 16, 18 times the
noise level which is $0.29~mJy~beam^{-1}$. The galaxy center is marked with a filled triangle.
The radio emission is mostly associated with the spiral arms and the southern arm is more prominent.
}
\end{figure}

\begin{figure}
\includegraphics[width=80mm,height=80mm,angle=-90]{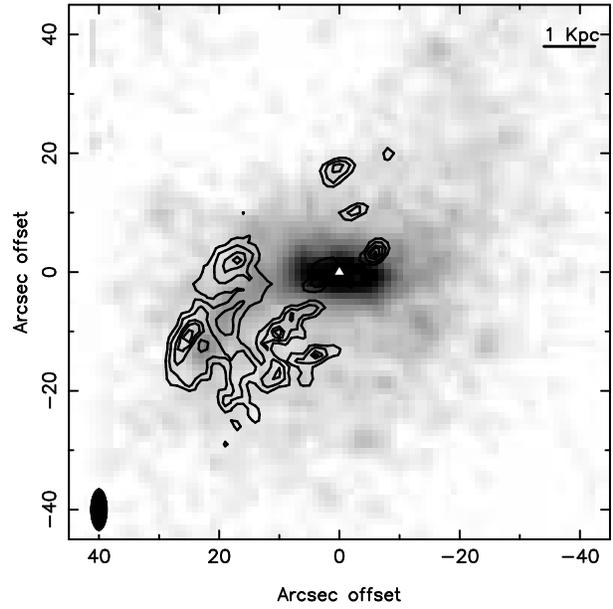}
\caption{Figure shows the contours of the uniform weighted (robust 0) GMRT 1280~MHz radio
continuum map of NGC~4027 superimposed on the 2MASS near-IR image of the galaxy. The peak radio flux
is $\sim1.9~mJy~beam^{-1}$ ($beam\sim5.7^{\prime\prime}$) and is located in the disk. 
The contours are 12, 14, 16, 17, 18 times the noise level which is $0.1~mJy~beam^{-1}$. 
Emission is mostly associated with knots of star formation in the southern spiral arm.
}
\end{figure}

\begin{figure}
\includegraphics[width=80mm,height=80mm,angle=-90]{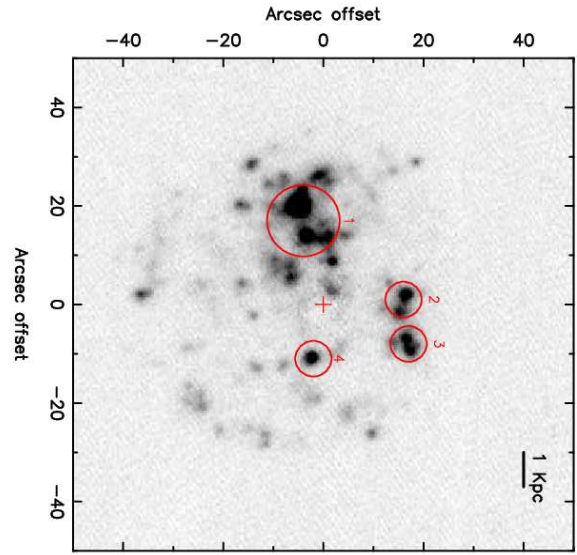}
\caption{The figure shows the H$\alpha$ image of NGC~3445 with the main star forming regions circled
in red (see Table~3). The emission is mainly concentrated along the southern spiral arm 
and  associated with star formation in the disk. The galaxy center is
marked with a red cross and does not seem to show much H$\alpha$ emission.}
\label{ha_n3445}
\end{figure}

\begin{figure}
\includegraphics[width=80mm,height=80mm,angle=-90]{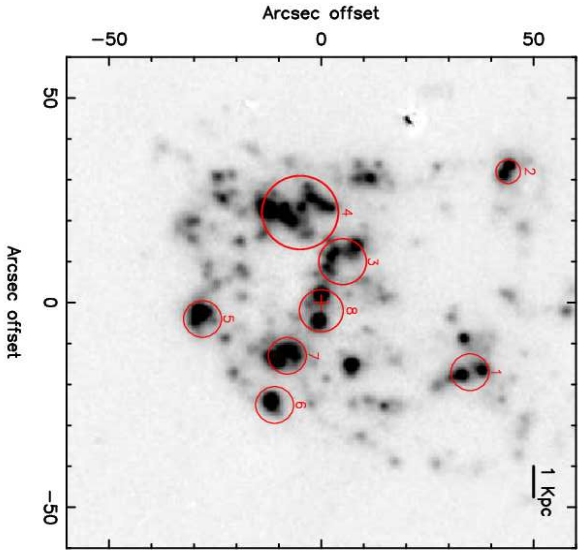}
\caption{The above figure shows the H$\alpha$ image of NGC~4027; the galaxy center is 
marked with a red cross and the main star forming regions circled in red (see Table~3). A large part 
of the emission is associated with the southern spiral arm and lies southeast of the galaxy nucleus. 
There are also star forming regions associated with the galaxy center and the northern spiral arm.} 
\label{ha_n4027}
\end{figure}

\begin{figure}
\includegraphics[width=80mm,height=80mm,angle=-90]{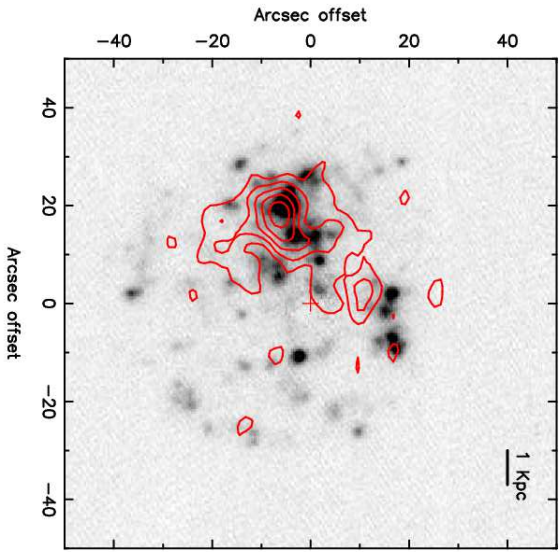}
\caption{The figure shows the H$\alpha$ image of NGC~3445 with radio continuum contours overlaid in 
red. The radio emission mostly traces the H$\alpha$ emission, but is more extened to the
south. }
\label{ha_rad_n3445}
\end{figure}

\begin{figure}
\includegraphics[width=80mm,height=80mm,angle=-90]{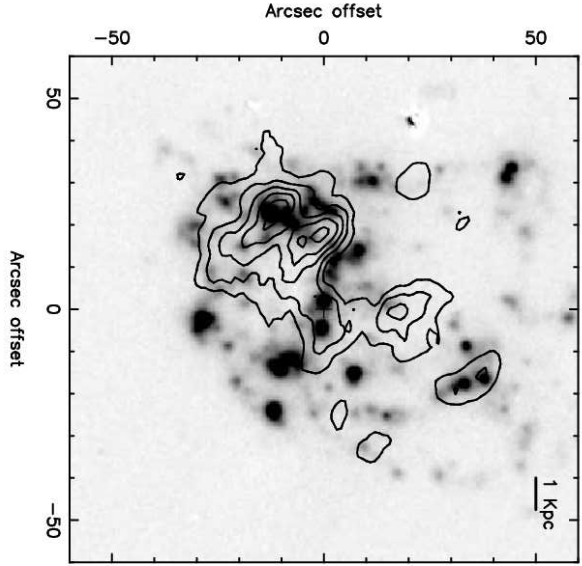}
\caption{The figure shows the H$\alpha$ image of NGC~4027 with radio continuum contours
overlaid in red. The radio emission follows the H$\alpha$ emission and traces the southern 
and northern spiral arms.}
\label{ha_rad_n4027}
\end{figure}

\begin{figure}
\includegraphics[width=80mm,height=70mm,angle=0]{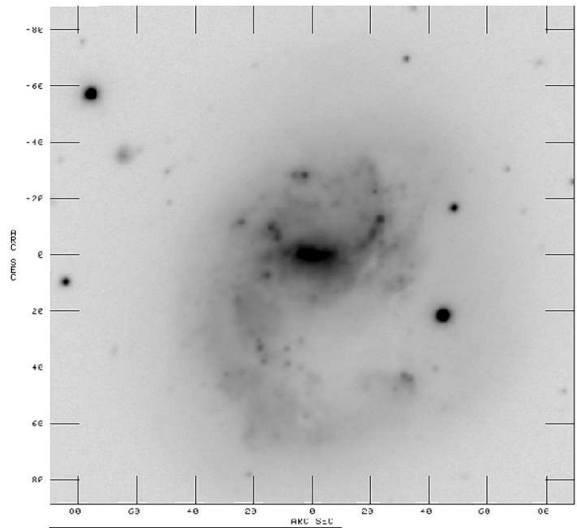}
\caption{Residual R band image of NGC~4027 derived from the H$\alpha$ image of the galaxy. The central oval distortion is
possibly a pseudobulge that developed from the buckling of a small bar triggered through interaction with NGC~4027~A.}
\end{figure}

\begin{figure}
\includegraphics[width=80mm,height=80mm,angle=0]{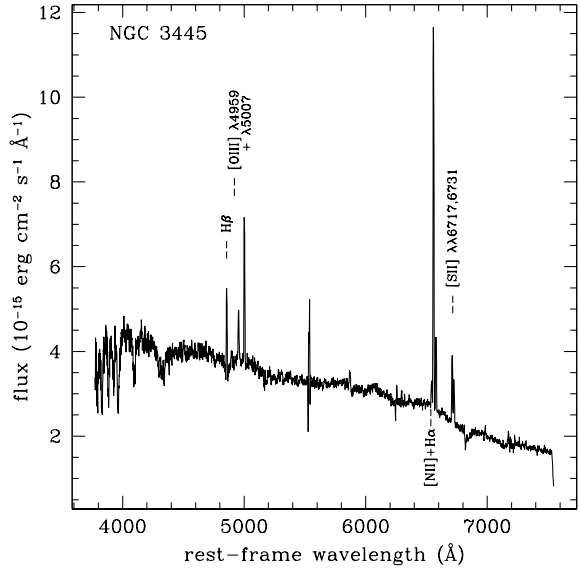}
\caption{Spectrum of NGC 3445 obtained from HFOSC installed in HCT. The spectrum covers the blue region obtained from 
grism \# 7 (3700-7800 \AA) having a dispersion of $\sim1.45$ \AA~pix$^{-1}$ at central wavelength 5600 \AA.}
\label{f:n3445_spec_r2}
\end{figure}

\begin{figure}
\includegraphics[width=80mm,height=80mm,angle=0]{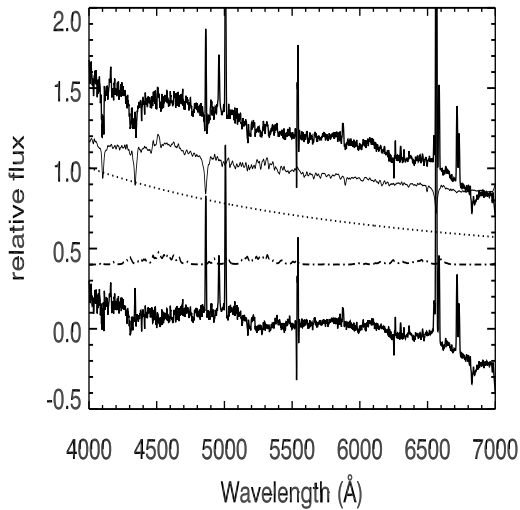}
\caption{The figure shows the observed spectrum (thick black line) of NGC 3445 and the best fitting starburst99 template 
of age $\sim700$ Myr (thin solid line). The dotted line is the power law template and the dot-dashed line is the 
Fe{\sc~ii} template. The difference between the observed spectrum and the combined model, power law and Fe{\sc~ii} 
templates. This residual spectrum is used for estimation of the integrated fluxes of each element.}
\label{f:n3445_sub_z_r2}
\end{figure}

\begin{figure}
\includegraphics[width=80mm,height=80mm,angle=0]{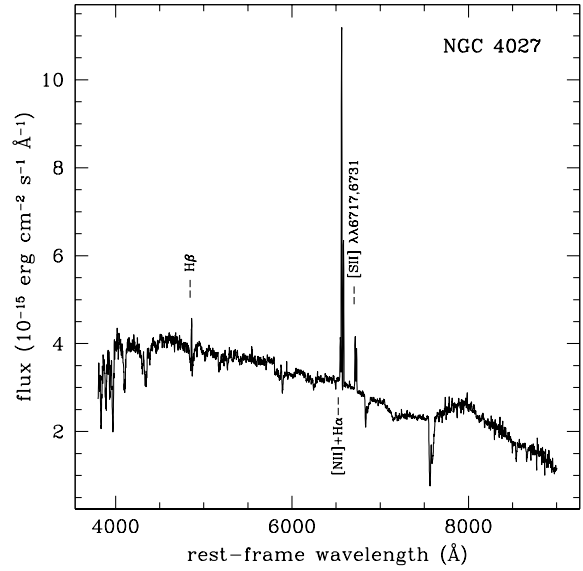}
\caption{Spectrum of NGC 4027 obtained from HFOSC installed in HCT. The spectrum is a combined spectrum of grism \#7 
and \#8 covering both the blue (3700-7800 \AA with dispersion of $\sim1.45$ \AA~pix$^{-1}$ at central wavelengths) and
 red region (5200-9100 \AA with a dispersion of $\sim1.25$ \AA~pix$^{-1}$ at central wavelengths). Notice the absence of 
Oxygen lines like \oii, \o3 or \oi.}
\label{f:n4027_spec_r2}
\end{figure}

\begin{figure}
\includegraphics[width=80mm,height=80mm,angle=0]{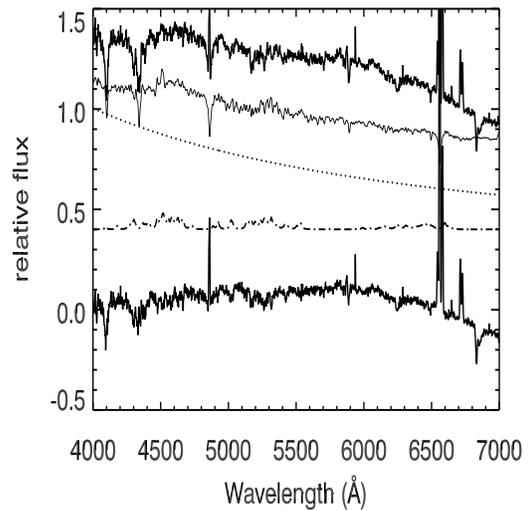}
\caption{Figure is similar to Figure \ref{f:n3445_sub_z_r2} but is for the galaxy NGC 4027. The best fit age for 
the underlying stellar population is $\sim1$ Gyr.}
\label{f:n4027_sub_z_r2}
\end{figure}

\end{document}